\documentclass[twocolumn]{aastex701}

\usepackage[T1]{fontenc} 
\usepackage[utf8]{inputenc} 
\usepackage{amssymb}
\usepackage{float}
\usepackage{placeins}
\usepackage{hyperref}
\usepackage{natbib}
\usepackage{lipsum}

\usepackage{subcaption}

\setcitestyle{notesep={ }}

\begin{document}

\title{POSEIDON II: The Anti-Aligned Orbit of the Warm Neptune TOI-1710 A b}

\correspondingauthor{Juan I.\ Espinoza-Retamal}
\email{jiespinozar@princeton.edu}

\author[0000-0001-9480-8526]{Juan I.\ Espinoza-Retamal}
\affiliation{Department of Astrophysical Sciences, Princeton University, 4 Ivy Lane, Princeton, NJ 08540, USA}
\email{jiespinozar@princeton.edu}

\author[0000-0002-5181-0463]{Hareesh Bhaskar} 
\affiliation{Department of Astronomy, Indiana University, Bloomington, IN 47405, USA}
\email{bhareeshg@gmail.com}

\author[0000-0002-4265-047X]{Joshua N.\ Winn}
\affiliation{Department of Astrophysical Sciences, Princeton University, 4 Ivy Lane, Princeton, NJ 08540, USA}
\email{jnwinn@princeton.edu}

\author[0000-0003-0412-9314]{Cristobal Petrovich}
\affiliation{Department of Astronomy, Indiana University, Bloomington, IN 47405, USA}
\email{cpetrovi@iu.edu}

\author[0000-0002-9158-7315]{Rafael Brahm}
\affil{Facultad de Ingenier\'ia y Ciencias, Universidad Adolfo Ib\'{a}\~{n}ez, Av. Diagonal las Torres 2640, 7941169 Pe\~{n}alol\'{e}n, Santiago, Chile}
\email{rafael.brahm@uai.cl}

\author[0000-0001-9985-0643]{Caleb Lammers}
\affiliation{Department of Astrophysical Sciences, Princeton University, 4 Ivy Lane, Princeton, NJ 08540, USA}
\email{caleb.lammers@princeton.edu}

\author[0000-0001-7409-5688]{Guðmundur Stefánsson}
\affil{Astrophysics \& Space Institute, Schmidt Sciences, New York, NY 10011, USA}
\affil{Anton Pannekoek Institute for Astronomy, University of Amsterdam, Science Park 904, 1098 XH Amsterdam, The Netherlands}
\email{g.k.stefansson@uva.nl}

\author[0009-0007-0740-0954]{Elise Koo}
\affil{Anton Pannekoek Institute for Astronomy, University of Amsterdam, Science Park 904, 1098 XH Amsterdam, The Netherlands}
\affil{ASTRON, Netherlands Institute for Radio Astronomy, Oude Hoogeveensedijk 4, Dwingeloo 7991 PD, The Netherlands}
\email{e.j.m.koo@uva.nl}

\author[0000-0002-5389-3944]{Andr\'es Jord\'an}
\affil{Facultad de Ingenier\'ia y Ciencias, Universidad Adolfo Ib\'{a}\~{n}ez, Av. Diagonal las Torres 2640, 7941169 Pe\~{n}alol\'{e}n, Santiago, Chile}
\affil{Departamento de Astronomía, Universidad de Chile, Casilla 36-D, Santiago, Chile}
\affil{El Sauce Observatory --- Obstech, Coquimbo, Chile}
\email{andres.jordan@uai.cl}

\author[0000-0003-3047-6272]{Felipe I. Rojas}
\affiliation{Instituto de Astrof\'isica, Pontificia Universidad Cat\'olica de Chile, Av. Vicu\~na Mackenna 4860, 7820436 Macul, Santiago, Chile}
\email{firojas@uc.cl}

\begin{abstract}

\added{We present an observation of the Rossiter-McLaughlin effect for the TOI-1710 system with the NEID spectrograph on the WIYN 3.5 m telescope. The system hosts a warm Neptune ($P\sim24$ days), and our observations reveal that it} orbits in the opposite direction to the stellar spin, with a sky-projected obliquity $\lambda=179\pm19^{\circ}$. Combined with information about the rotation period of the host star, we measure a true obliquity $\psi=158_{-13}^{+11}\,^{\circ}$. The host star has an M-dwarf companion at a separation of $\sim3600$~au, but this companion is too distant to be solely responsible for misaligning the warm Neptune. The host star also shows a long-term radial velocity trend, indicative of a companion at intermediate
separations. We show that such a companion can dynamically couple the warm Neptune to the distant M dwarf, enabling the transfer of inclination from the wide binary orbit to the planetary orbit. Assuming this scenario is correct, we predict the intermediate companion is a $\sim5\,M_J$ planet on a $\sim15$~au orbit that is nearly aligned with the transiting planet's orbit.

\end{abstract}


\section{Introduction} 

Stellar obliquities, the angles between stellar spin and planetary orbits, are probes of the dynamical histories of planetary systems \citep[e.g.,][]{Albrecht2022}. Measurements based on the Rossiter-McLaughlin (RM) effect \citep{Rossiter1924,McLaughlin1924} have revealed a wide diversity of system architectures, ranging from well-aligned to strongly misaligned and even retrograde configurations \citep[e.g.,][]{Hjorth2021,Wang2024,Espinoza-Retamal2025,Zak2025,Bourrier2022,Rubenzahl2024b}. Misalignments are commonly interpreted as signatures of dynamical processes such as high-eccentricity migration, yet their origin remains debated, particularly for lower-mass planets.

Most obliquity measurements have focused on short-period giant planets because they are observationally favorable, but they may not be representative of smaller and more commonly occurring exoplanets \citep[e.g.,][]{Howard2010,Batalha2013}. \added{Smaller planets may undergo distinct dynamical evolution. In addition, smaller planets are expected to be less efficient at tidally realigning their host stars than hot Jupiters, allowing their observed obliquities to serve as more reliable probes of their dynamical histories \citep[e.g.,][]{Zahn1977,Albrecht2012}.} Recent advances in radial velocity (RV) precision have begun to allow obliquity studies to be extended to Neptune-sized planets \citep[e.g.,][]{Stefansson2022,Bourrier2023,Bourrier2025,Handley2025,Yee2025,Tamburo2025,Polanski2025}. Expanding the sample of obliquity measurements in this regime is essential to assess whether the trends observed for giant planets also apply to the more common population of smaller planets.

The POSEIDON survey aims to characterize the obliquity distribution of transiting Neptunes and use it to constrain their dynamical origins \citep{Espinoza-Retamal2026}. As part of this effort, here we present observations of the RM effect and an obliquity measurement for the warm-Neptune system TOI-1710 \citep{Konig2022}. TOI-1710 is a wide binary composed of a G-type primary star and an M-dwarf companion separated by $\sim3600$~au \citep{El-Badry2021}. A transiting Neptune around the primary was identified by \citet{Konig2022}, and its properties were refined by subsequent analyses \citep{Orell-Miquel2024,Polanski2024}. The transiting
planet has an orbital period of $\sim24$~d, a Neptune-like mass of $\sim20\,M_\oplus$, and a radius of $\sim5\,R_\oplus$, placing the planet in the ``Neptune savanna'' \citep{Bourrier2023}.

\section{Observations}\label{sec:observations}

\subsection{NEID Transit Spectroscopy}

We observed one transit of TOI-1710 A b using the NEID spectrograph \citep{Schwab2016} installed on the 3.5 m WIYN telescope at Kitt Peak Observatory in Arizona. NEID is an echelle spectrograph that covers the wavelength range of $380-930$ nm at a resolving power of $R\approx110,\!000$. The transit was observed on UTC October 25, 2025, between 05:39 and 12:26. We obtained 37 spectra spanning the transit with exposure times of 600 s. The spectra were processed with the NEID Data Reduction Pipeline version 1.4.2 \citep{Bender2022}, which resulted in final spectra with a median signal-to-noise ratio of 68 per pixel at 550 nm. We extracted precise RVs using the \texttt{serval} template-matching code \citep{Zechmeister2018} adapted for NEID by \citet{Stefansson2022}. The resulting RVs have a median uncertainty of 1.0 m s$^{-1}$, derived from the order-by-order uncertainties. The NEID RVs are shown in Figure~\ref{fig:fit_1710} and are tabulated in Appendix~\ref{app:rvs}.

\begin{figure*}
    \centering
    \includegraphics[width=0.83\linewidth]{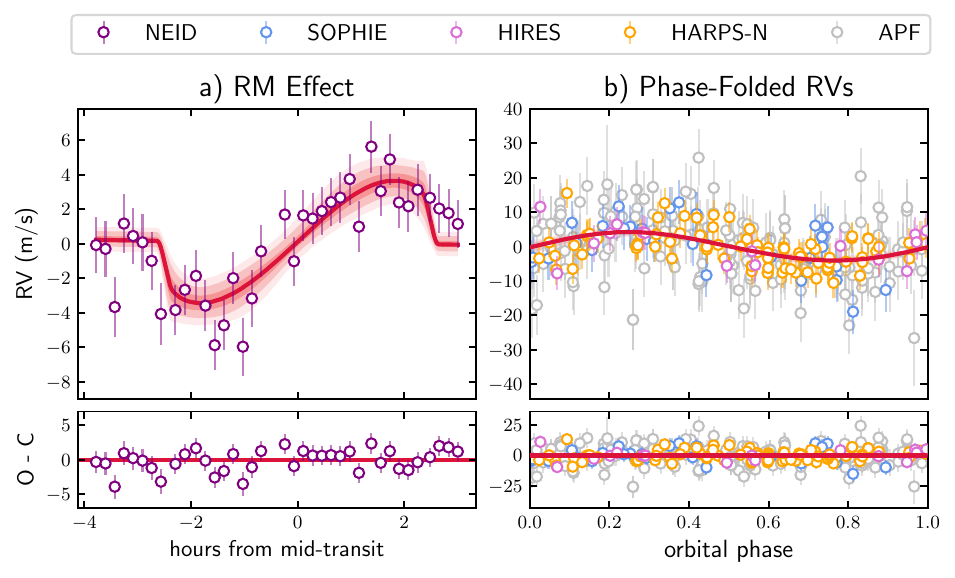}
    \includegraphics[width=0.83\linewidth]{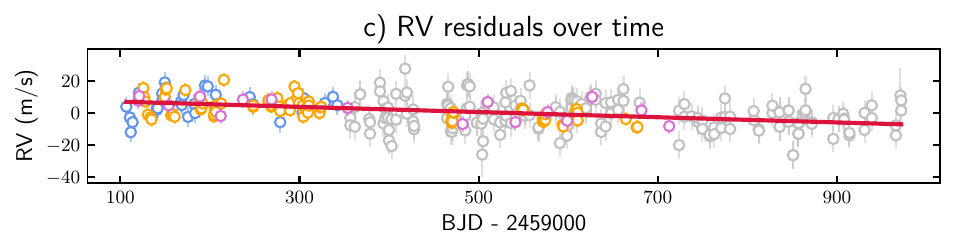}
    \includegraphics[width=0.83\linewidth]{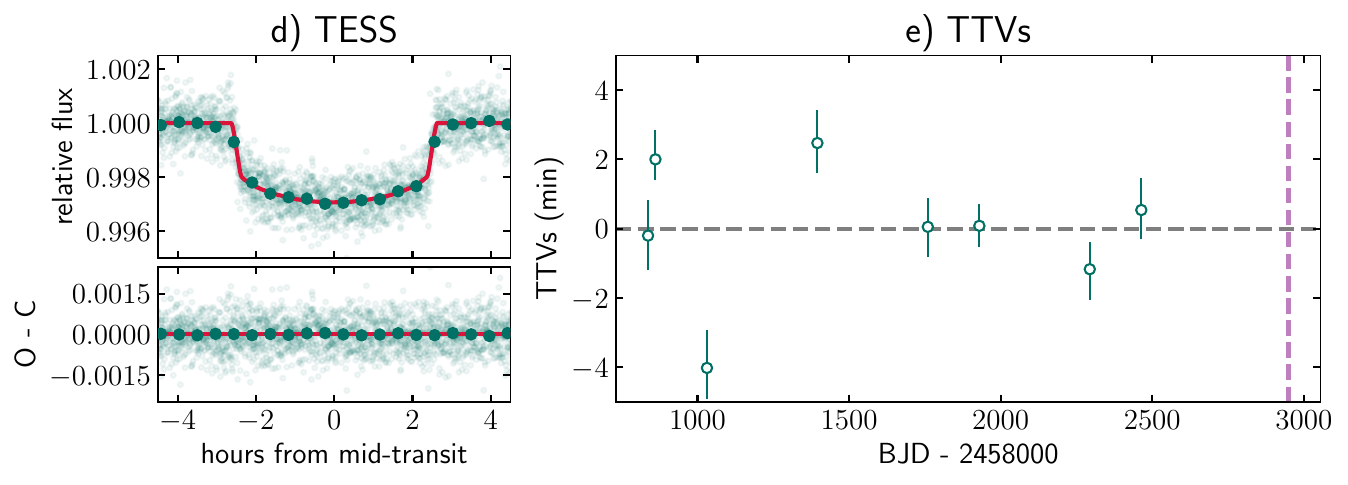}
    \caption{Observations of TOI-1710~A. (a) NEID RVs, after subtracting a long-term linear trend (purple) along with the best-fit model including the RM effect
    (red curve) and the associated confidence intervals (1, 2, and $3\sigma$, shaded red). Residuals are shown below. (b) Out-of-transit RVs versus orbital phase, along with the best-fit model (the confidence intervals are too small to be
    seen clearly). Residuals are shown below. (c) RV residuals versus time, after subtracting the signal of TOI-1710~A~b. The red line is a linear RV trend, which is detected with 8$\sigma$ confidence. (d) Transit photometry from TESS (green) along with the best-fit model (red). The darker points are time-averaged data. Residuals are shown below. (e) Deviations between measured transit times and the best-fit constant period model as a function of time. The vertical dashed line indicates the transit that was observed spectroscopically with NEID.}
    \label{fig:fit_1710}
\end{figure*}

\subsection{TESS Photometry}

We retrieved the TESS light curves for Sectors 19, 20, 26, 40, 53, 59, 60, 73, and 79 from the Mikulski Archive for Space Telescopes using the \texttt{lightkurve} package \citep{lightkurve}. We opted for the light curves with 120~s sampling as processed by the Science Processing Operations Center pipeline \citep{spoc}, which corrects for pointing and focus-related instrumental signatures, discontinuities resulting from radiation events in the CCD detectors, outliers, and contributions to the recorded flux from nearby stars. The TESS transit observations, along with the best model, are shown in Figure~\ref{fig:fit_1710}.

\subsection{Archival Spectroscopy}

To determine the spectroscopic parameters of the orbit, we used archival RV observations: 30 from SOPHIE \citep{Konig2022}, 85 from HARPS-N \citep{Orell-Miquel2024}, 178 from the APF, and 16 from HIRES \citep{Polanski2024}. We also used archival Keck/HIRES spectra\footnote{\url{https://koa.ipac.caltech.edu/cgi-bin/KOA/nph-KOAlogin}} and processed them with the \texttt{ceres} pipeline \citep{ceres} for stellar spectral characterization. 

\section{Stellar Parameters}\label{sec:stellar}

We derived the stellar parameters of TOI-1710~A following the iterative procedure described by \citet{Brahm2019}. First, we determined the atmospheric parameters from the HIRES spectrum using \texttt{zaspe} \citep{zaspe}. We then estimated the physical parameters by fitting the star's spectral energy distribution
(based on broadband photometry) to synthetic magnitudes from the \texttt{PARSEC} isochrones \citep{parsec} and the Gaia DR3 parallax \citep{GaiaDR3}. The resulting mass and radius were used to calculate $\log g$, which was then held fixed in a new iteration of the spectroscopic analysis. We repeated this process until convergence in $\log g$. The final stellar parameters are listed in Appendix~\ref{app:stellar}.

\section{Photometric Analysis}\label{sec:phot}

In order to refine the orbital ephemeris and to look for possible transit timing variations (TTVs), we analyzed the TESS data with the \texttt{juliet} code \citep{juliet}. We modeled the transits with \texttt{batman} \citep{batman} and included a quasi-periodic (QP) Gaussian Process (GP) with \texttt{celerite} \citep{celerite} to remove periodic modulation related to the rotational period of TOI-1710~A. We placed broad Gaussian priors on each transit midpoint (width of 1~day) based on the ephemeris of \citet{Konig2022}.

As shown in Figure~\ref{fig:fit_1710}, the transit times from Sectors 20, 26, and 40 deviate from the best-fit strictly periodic model by approximately 2.0, 4.0, and 2.5 minutes, respectively. The associated statistical significances are $2.4\sigma$, $3.6\sigma$, and $2.8\sigma$. Although formally significant, we regard the evidence for TTVs as suggestive, not conclusive. The TESS light curves exhibit structured, time-correlated noise that may have caused the timing uncertainties to be underestimated. Additional transit observations will be valuable for determining whether TOI-1710~A~b exhibits TTVs related to the presence of companion planets. From the TTV analysis, we obtained an improved orbital ephemeris and a detrended TESS light curve, which we used in the analysis described below.

Finally, we note that the posterior for the QP-GP period, $26\pm16$~d, is consistent with previous estimates of the stellar rotation period. \citet{Konig2022} reported $P_{\rm rot}=22.5\pm2.0$~d based on periodic signals identified in multiple stellar activity indicators (e.g., chromospheric emission indices) and calibrated activity-rotation relations. Consistently, \citet{Orell-Miquel2024} also found evidence for variability on similar timescales ($\sim20$--$24$~d) in ground-based photometry and modeled the RVs using a QP-GP with a characteristic timescale consistent with stellar rotation.

\begin{deluxetable*}{llcr}[h!]
\tablecaption{Summary of priors and posteriors of the \texttt{ironman} fit. The subscript ``A'' refers to the primary star, and
``b'' refers to the transiting planet.\label{tab:fit_1710}}
\tablewidth{70pt}
\tablehead{Parameter & Description & Prior & Posterior}
\startdata
$\psi$ & True stellar obliquity ($^{\circ}$) & \nodata & $158_{-13}^{+11}$ \\
$\lambda$ & Sky-projected stellar obliquity ($^{\circ}$) & $\mathcal{U}(0,360)$ & $179\pm19$ \\
$v$ & Equatorial velocity (km s$^{-1}$) & \nodata & $2.2\pm0.2$ \\
$v\sin{i_A}$ & Projected rotational velocity (km s$^{-1}$) & \nodata & $2.1\pm0.2$ \\
$\cos{i_A}$ & Cosine of stellar inclination & $\mathcal{U}(-1,1)$ & $0.0\pm0.3$ \\
$i_A$ & Stellar inclination ($^{\circ}$) & \nodata & $91_{-19}^{+18}$ \\
$P_{\rm rot}$ & Stellar rotational period (d) & $\mathcal{G}(22.5,2.0)$ & $21.3\pm1.8$ \\
$M_A$ & Stellar mass ($M_\odot$) & $\mathcal{G}(1.023,0.015)$ & $1.020\pm0.015$ \\
$R_A$ & Stellar radius ($R_\odot$) & $\mathcal{G}(0.93,0.01)$ & $0.93\pm0.01$ \\
$\rho_A$ & Stellar mean density (g cm$^{-3}$) & \nodata & $1.77\pm0.06$ \\
\hline
$P$ & Orbital period (d) & $\mathcal{U}(24.28327,24.28346)$ & $24.28336\pm0.00001$ \\ 
$t_0$ & Transit midpoint (BJD) & $\mathcal{U}(2458812.6757,2458812.6833)$ & $2458812.6797\pm0.0004$ \\
$b$ & Impact parameter & $\mathcal{U}(-1,1)$ & $0.00\pm0.12$ \\
$i_b$ & Orbital inclination ($^{\circ}$) & \nodata & $90.01\pm0.18$ \\
$R_b/R_A$ & Radius ratio & $\mathcal{U}(0,1)$ & $0.0497\pm0.0003$ \\
$K$ & RV semiamplitude (m s$^{-1}$) & $\mathcal{U}(0,1000)$ & $4.2\pm0.5$ \\
$\sqrt{e}\sin{\omega}$ & Eccentricity parameter -- sine component & $\mathcal{U}(-1,1)$ & $-0.15\pm0.05$ \\
$\sqrt{e}\cos{\omega}$ & Eccentricity parameter -- cosine component & $\mathcal{U}(-1,1)$ & $0.09_{-0.20}^{+0.15}$ \\
$e_b$ & Eccentricity & \nodata & $0.05_{-0.02}^{+0.04}$ \\
$\omega_b$ & Argument of periastron ($^{\circ}$) & \nodata & $-62_{-64}^{+36}$ \\
$a_b/R_A$ & Scaled semimajor axis & \nodata & $38.0_{-0.5}^{+0.4}$ \\
$R_b$ & Planet radius ($R_\oplus$)& \nodata & $5.06\pm0.06$ \\
$a_b$ & Semimajor axis (au)& \nodata & $0.1652\pm0.0008$ \\
$M_b$ & Planet mass ($M_\oplus$)& \nodata & $19.1\pm2.3$ \\
$\rho_b$ & Planet density (g cm$^{-3}$)& \nodata & $0.8\pm0.1$ \\
$\dot{\gamma}$ & RV slope (m s$^{-1}$ d$^{-1}$)& $\mathcal{U}(-1,1)$ & $-0.016\pm0.002$ \\
\hline
$q_1^{\rm NEID}$ & NEID linear limb darkening parameter & $\mathcal{U}(0,1)$ & $0.53\pm0.29$ \\
$q_2^{\rm NEID}$ & NEID quadratic limb darkening parameter & $\mathcal{U}(0,1)$ & $0.57_{-0.34}^{+0.29}$ \\
$\beta$ & Intrinsic stellar line width (km s$^{-1}$) & $\mathcal{G}(4.7,2.0)$ & $3.5_{-2.0}^{+2.4}$ \\
$\gamma_{\rm NEID}$ & NEID RV offset (m s$^{-1}$)& $\mathcal{U}(-500,500)$ & $24\pm3$ \\
$\sigma_{\rm NEID}$ & NEID RV jitter (m s$^{-1}$)& $\mathcal{LU}(10^{-3},100)$ & $1.1\pm0.3$ \\
$\gamma_{\rm SOPHIE}$ & SOPHIE RV offset (m s$^{-1}$)& $\mathcal{U}(-39000,-38000)$ & $-38860.1_{-1.3}^{+1.4}$ \\
$\sigma_{\rm SOPHIE}$ & SOPHIE RV jitter (m s$^{-1}$)& $\mathcal{LU}(10^{-3},100)$ & $5.9_{-0.8}^{+1.0}$ \\
$\gamma_{\rm HIRES}$ & HIRES RV offset (m s$^{-1}$)& $\mathcal{U}(-500,500)$ & $-3.2\pm1.3$ \\
$\sigma_{\rm HIRES}$ & HIRES RV jitter (m s$^{-1}$)& $\mathcal{LU}(10^{-3},100)$ & $4.6_{-0.9}^{+1.2}$ \\
$\gamma_{\rm HARPS-N}$ & HARPS-N RV offset (m s$^{-1}$)& $\mathcal{U}(-500,500)$ & $-4.9\pm0.6$ \\
$\sigma_{\rm HARPS-N}$ & HARPS-N RV jitter (m s$^{-1}$)& $\mathcal{LU}(10^{-3},100)$ & $4.1_{-0.3}^{+0.4}$ \\
$\gamma_{\rm APF}$ & APF RV offset (m s$^{-1}$)& $\mathcal{U}(-500,500)$ & $1.1\pm0.6$ \\
$\sigma_{\rm APF}$ & APF RV jitter (m s$^{-1}$)& $\mathcal{LU}(10^{-3},100)$ & $7.9_{-0.4}^{+0.5}$ \\
\hline
$q_1^{\rm TESS}$ & TESS linear limb darkening parameter & $\mathcal{U}(0,1)$ & $0.20_{-0.06}^{+0.08}$ \\
$q_2^{\rm TESS}$ & TESS quadratic limb darkening parameter & $\mathcal{U}(0,1)$ & $0.60_{-0.17}^{+0.2}$ \\
$\sigma_{\rm TESS}$ & TESS photometric jitter (ppm) & $\mathcal{LU}(1,5\times10^7)$ & $94_{-77}^{+33}$ \\
\enddata
\tablecomments{$\mathcal{U}(a,b)$ denotes a uniform prior with a start value $a$ and end value $b$. $\mathcal{G}(\mu,\sigma)$ denotes a normal prior with mean $\mu$, and standard deviation $\sigma$. $\mathcal{LU}(a,b)$ denotes a log-uniform prior with a start value $a$ and end value $b$.}
\end{deluxetable*}

\section{Obliquity Determination}\label{sec:fit}

To constrain the stellar obliquity, we performed a joint analysis of the observations described in Section \ref{sec:observations} using the \texttt{ironman}\footnote{\url{https://github.com/jiespinozar/ironman}} package \citep{Espinoza-Retamal2023b,Espinoza-Retamal2024}. \texttt{ironman} is a versatile \texttt{Python} package that can jointly fit the RM effect, Keplerian RVs, and transit photometry, by interfacing with other packages: \texttt{rmfit} \citep{Stefansson2022} for the RM effect modeling, \texttt{batman} \citep{batman} for the transit light curves, \texttt{radvel} \citep{Fulton18} for the Keplerian RVs, and \texttt{dynesty} \citep{dynesty2} to sample the posteriors.

We adopt the stellar rotation period reported by \citet{Konig2022} to derive the true obliquity using the parametrization from \citet{Stefansson2022} available in \texttt{ironman}. This parametrization performs the correct accounting for the correlation between $v\sin{i}$ and the equatorial velocity of the star \citep[see][]{Masuda2020}. In this analysis we worked with the TESS light curve that was detrended in Section~\ref{sec:phot}, and considered only within 10 hours of the transits to reduce computational cost. We included independent jitter terms for each photometric and RV instrument to account for possible systematics. We placed uniform priors on almost all parameters, except for the stellar parameters (mass, radius, rotation period). All priors and results obtained from the posterior distributions are shown in Table~\ref{tab:fit_1710}.

We find that TOI-1710~A~b has a nearly circular ($e=0.05_{-0.02}^{+0.04}$) and retrograde orbit with $\lambda=179\pm19\,^{\circ}$ and $\psi=158_{-13}^{+11}\,^{\circ}$. We also noticed the presence of a long-term RV trend (see Figure~\ref{fig:fit_1710}). We modeled it in our joint fit using a linear function of the form $\dot{\gamma}\left(t-t_a\right)$ where we arbitrarily chose $t_a=2459539.2192$, near the middle of the observing baseline. We compared the models with and without the linear trend using the Bayesian evidence ($\log{Z}$). The model with the linear trend is favored by $\Delta\log{Z}=18$. The RV slope is $\dot{\gamma}=-0.016\pm0.002$ m s$^{-1}$ d$^{-1}$, an $8\sigma$ detection. Overall, the parameters of the system agree to within $2\sigma$ with those reported by \citet{Konig2022}, \citet{Orell-Miquel2024}, and \citet{Polanski2024}.

\section{Discussion}

\subsection{Search for Companions}\label{sec:companions}

The catalog of \citet{El-Badry2021} includes a spatially resolved M-dwarf companion to TOI-1710\,A, identified based on similar proper motion and parallax. The companion has Gaia DR3 identifier 1116612783096856960, and will hereafter be called TOI-1710~B. The current projected separation is $\approx45^{\prime\prime}$, which translates into $\approx3610$ au.
Although the Gaia parallaxes of the two stars are consistent,
their proper motions differ by a small but significant amount, and this relative sky-plane velocity can be used to constrain the binary orbit. For this purpose, we used the \texttt{lofti\_gaia} package \citep{Pearce2020}, which constrains binary orbits using Gaia astrometry by fitting the relative position and proper motion of the system at a single epoch. In brief, the code generates trial Keplerian orbits, scales and rotates them to match the observed projected separation, and then accepts or rejects them based on how well they reproduce the observed relative velocity. The code requires as inputs the Gaia IDs of both stars and knowledge of their masses. For TOI-1710~A, we used the value reported in Appendix~\ref{app:stellar} of $1.023\pm0.015\,M_{\odot}$, while for TOI-1710~B, we used $0.40\pm0.02\,M_{\odot}$, the value reported in the TESS Input Catalog \citep{Stassun2018,Stassun2019}. After 100,000 accepted orbits we obtained a semimajor axis $a_B=3354^{+1056}_{-1527}$ au, inclination $i_B=103_{-8}^{+7}\,^{\circ}$, and eccentricity $e_B=0.82^{+0.16}_{-0.15}$. The reported parameters for the orbit of TOI-1710~B are the median and 68\% credible intervals of the accepted orbits. As the planet's orbit has
$i_b\approx 90^\circ$, the minimum angle between the planetary and binary orbital planes is $\approx13\pm8^{\circ}$.

We also evaluate the possibility of additional planetary companions. As discussed in Section~\ref{sec:phot}, there is suggestive but inconclusive evidence for TTVs. A search for additional transiting planets using a box least squares algorithm did not reveal any significant signals. In contrast, the RVs show evidence for a long-term trend. This trend cannot be attributed to the wide binary companion, as its expected contribution based on the derived orbital parameters is several orders of magnitude too small. If instead the RV trend is produced by a planetary companion, the planet's orbital period must be longer than the current RV baseline ($\gtrsim 2.4$ yr). The system is therefore a promising target for future astrometric constraints with Gaia \citep[e.g.,][]{Perryman2014,Espinoza-Retamal2023a,Lammers2025}, given its proximity and brightness ($d\approx81$ pc, $G \approx 9.4$ mag). Unfortunately, it was not observed by Hipparcos, and is therefore not included in the Hipparcos-Gaia catalog of accelerating sources \citep{Brandt2021}, limiting current astrometric constraints on additional companions. No additional periodic signals are detected in the RVs after removing the signal of TOI-1710~A~b and the long-term trend.

\begin{figure*}
    \centering
    \includegraphics[width=\linewidth]{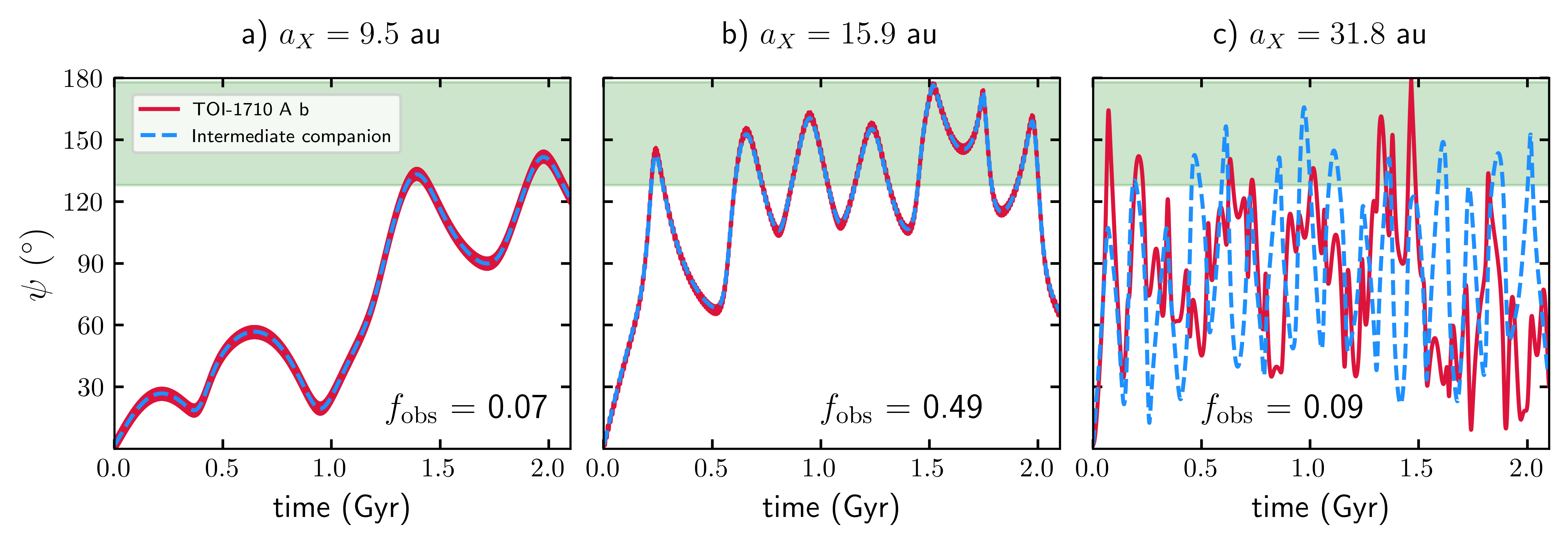}
    \caption{Secular evolution of the obliquities of TOI-1710 from four-body secular integrations. The semimajor axis of the intermediate companion varies across panels. The obliquity evolution of TOI-1710~A~b (red) and the hypothetical intermediate companion X (blue) are shown. The companion mass is fixed at $9\,M_J$ and the initial mutual inclination with the M-dwarf is $i_{XB} = 55^\circ$. The green region marks the $2\sigma$ constraints on the obliquity of the observed planet, with the fraction of time spent within this region indicated by $f_{\rm obs}$. Panels (a) and (c) remain within the allowed window for much less time than panel (b). In all cases, the eccentricity of planet b remains nearly the same as the initial value.}
    \label{fig:trajobq}
\end{figure*}

\subsection{Possible Origins of the Retrograde Orbit}\label{sec:origins}

How can we explain the retrograde and nearly circular orbit of TOI-1710~A~b? 

\paragraph{Disk torquing}  One possibility is that the protoplanetary disk was strongly tilted with respect to the stellar spin axis. Such disk misalignments might be caused by gravitational torques from stellar companions \citep[e.g.,][]{Batygin2012,Lai2014,Zanazzi2018}. However, the nodal precession timescale of a disk around star A induced by the torque from star B is likely too long compared to the disk dispersal timescale to efficiently excite large stellar obliquities. Using the binary properties described in Section~\ref{sec:companions} and following \citet{Lai2014}, we estimate that a protoplanetary disk extending out to 100~au would complete half a precession cycle in $\sim40$~Myr, significantly exceeding typical protoplanetary disk lifetimes \citep[e.g.,][]{Ercolano2017}. Alternatively, the stellar obliquity may have been excited through magnetic disk-star interactions during the early accretion phase, in which torques exerted by a magnetically coupled disk can tilt the stellar spin axis \citep[e.g.,][]{Lai2011,Spalding2015}. 

\paragraph{High-eccentricity tidal migration} 
\added{Secular perturbations from the distant stellar companion or an unseen massive planetary companion might have driven high-eccentricity migration (HEM) of the Neptune-mass planet, possibly
explaining the retrograde orbit of TOI-1710~A~b. However, this scenario requires efficient tidal dissipation to circularize the orbit. This seems unlikely, because the expected
circularization timescale greatly exceeds the age of the system.\footnote{Following \citet{Goldreich66} and \citet{Hut81}, and adopting a modified tidal quality factor \citep[e.g.,][]{Ogilvie07} of $Q^\prime = 10^4$, we estimate a tidal circularization timescale of $\sim 30$ Gyr.} 
Based on Equation~7 of \cite{Owen2018},
the Neptune's final semimajor axis $a_{\rm circ}$ resulting from HEM must satisfy
\begin{eqnarray}
&& a_{\rm circ}\lesssim 0.16 \,{\rm au}\, \left({t_{\rm age}\over {\rm1\, Gyr}}\right)^{1/7}
\left({Q_b'\over 10}\right)^{-1/7}\left({M_\mathrm{A}\over M_\odot}\right)^{2/7}\nonumber\\
&& \qquad\times \left({a_0\over {\rm au}}\right)^{-1/7}
\left({M_b\over 19 M_\oplus}\right)^{-1/7}
\left({R_b\over 5 R_\oplus}\right)^{5/7}
\label{eq:amig}
\end{eqnarray}
 where $a_0$ is the initial semimajor axis and
 $Q_b'$ is the ratio of planet's tidal quality factor $Q$ and its Love number $k_{2}$. Thus, obtaining $a_{\rm circ} = 0.16$~au
 requires either $Q_b'\lesssim 10$,
 which seems unrealistically small,
 or a planetary radius substantially
 larger than the current value of $5\,R_\oplus$.
 In some other systems, internal heat from
 tidal dissipation is likely to have
 slowed the planet's contraction
 and allowed it to remain large
 over HEM timescales \citep{roznerInflatedEccentricMigration2022,yuAreWASP107likeSystems2024,luPlanetPlanetScatteringZLK2024a,hallattSheddingLightDesert2025}.
 In this case, though, 
 the expected internal heat flux
 from tidal dissipation ($F_\mathrm{tide}\sim1-100\,\mathrm{erg\,cm^{-2}\,s^{-1}}$ for a time lag of 10~s) is much smaller than the
 external flux from stellar irradiation ($F_\mathrm{star}\sim10^6-10^8\,\mathrm{erg\,cm^{-2}\,s^{-1}}$)
 over the planet's entire migration track. Therefore, if the planet was inflated,
 it was probably due to stellar irradiation
 rather than tidal dissipation. While the dependence on $Q'$ is weak, the orbit of a sufficiently inflated planet could have been circularized. While this is an interesting possibility, it is not considered here for simplicity.}

\added{\paragraph{Roche lobe overflow}
A giant planet might undergo Roche lobe overflow if it migrates too close to its
host star. Tidal stripping of the gaseous
envelope would leave behind a planet with a significantly reduced mass, smaller radius, and wider final orbit. 
However, the orbit
of TOI-1710~A~b is too wide for
the planet to be the tidally stripped
remnant of a giant planet. \cite{valsecchiTIDALLYDRIVENROCHELOBE2015} found that giant planets with core masses of $\sim 15M_\oplus$, comparable to that of TOI-1710~A~b, typically end up on orbits with periods of only $\sim1$~day.}

\paragraph{Secular inclination cascade driven by the distant star} 
For the rest of this section, we consider a scenario in which the orbit of TOI-1710~A~b is tilted through secular interactions with the M-dwarf companion and the additional substellar companion that produces the long-term radial velocity trend. Such a companion, hereafter called planet X, could facilitate the transfer of inclination from the binary orbit to the planetary orbit in a cascading\footnote{\added{We borrow the term from \citet{Yang2025}, who considered an ``eccentricity cascade'' by which angular-momentum deficit is transferred from a very wide companion to a cold Jupiter through an intermediate-separation companion, eventually triggering tidal migration. In the case considered here, the companion enables the transfer of inclination, a scenario that we name the ``inclination cascade''.}} manner. This mechanism is similar to that proposed by \citet{Best2022} for the retrograde system K2-290 \citep{Hjorth2019,Hjorth2021}.

\subsubsection{Numerical Experiments of the Secular Inclination Cascade Driven by the Distant Companion} 

We adopt the currently observed orbital configuration of TOI-1710~A~b as the initial condition, namely a semimajor axis of $a=0.16$~au and a nearly circular orbit ($e = 0.05^{+0.04}_{-0.02}$). \added{This could be the result of in-situ formation or disk migration}. We also adopt the present-day stellar rotation period of $21.3 \pm 1.8$~days, because obliquity evolution occurs on timescales of hundreds of Myr after formation, over which the stellar spin is not expected to vary significantly.

The known M-dwarf companion, with a semimajor axis of $3354^{+1056}_{-1527}$~au, is too distant to overcome
the planet's strong coupling to the stellar spin,
and therefore cannot by itself
generate the extreme obliquity of TOI-1710~A~b ($\psi = 158^{+11}_{-13}\,\mathrm{deg}$).\footnote{The Laplace radius associated with the M-dwarf companion is 0.6~au, placing TOI-1710~A~b well inside the regime where spin--orbit coupling is dominant.} We therefore construct a hierarchical four-body model comprising the host star, planet~b, planet~X, and the distant M-dwarf companion, and perform an ensemble of secular simulations to constrain the mass and semimajor axis of planet~X (see Appendix~\ref{app:simulations} for details).

Figure~\ref{fig:trajobq} shows three calculations of the secular evolution of the obliquities of planets b and X (i.e., their orbital
inclinations relative to the star's equatorial plane)
for different choices of planet X's semimajor axis ($a_X$).
In all cases, even though the obliquities are initially very
small, planet~b's obliquity evolves to reach the $2\sigma$ lower limit of $\psi = 140^\circ$, consistent with observations, while its eccentricity remains close to the initial value. To quantify the agreement with observations, we define $f_{\rm obs}$ as the fraction of the integration time during which the simulated system satisfies the obliquity constraints.

Panels~(a) and~(b) show simulations in which planets~b and~X are tightly coupled, allowing them to maintain nearly aligned orbital planes throughout the evolution. In panel~(a), the companion lies too close to the host star, leading to delayed obliquity excitation and only brief excursions into the observationally allowed range within the system’s age (low $f_{\rm obs}$). In contrast, panel~(b) shows a configuration in which the separation of planet~X is close to optimal for obliquity excitation; in this case, the system spends nearly half of the integration time within the observed obliquity range. In neither case can the evolution be explained
solely by nodal precession of the planetary orbits, for which
the maximum $\psi$ would be $2\,i_{\rm XB} \sim 110^\circ$. Instead, a secular spin--orbit resonance occurs at $a_X \sim 10$~au (Appendix~\ref{app:simulations}), where the stellar spin precession rate becomes comparable to the nodal precession rate induced by the M-dwarf companion. This resonance drives the evolution of the stellar spin vector, enabling the obliquity of planet~b to reach the observationally allowed range.

Panel~(c) illustrates a case in which a secular orbit--orbit resonance arises between the nodal precession of planet~X induced by the M-dwarf and that of planet~b induced by planet~X, leading to chaotic evolution. As in panel~(a), the system satisfies the observational constraints for only a small fraction of the integration time ($f_{\rm obs} < 0.1$).

\subsection{Predictions for Planet X}

Having shown that secular inclination excitation can account for both the low eccentricity and high obliquity of the observed
planet~b, we now combine the radial velocity trend with population-synthesis predictions to further constrain the properties of the hypothesized planet X.

The outcome of secular evolution depends sensitively on the mutual inclination between the planetary system and the stellar companion~B. We perform an ensemble of secular simulations
over a range of choices for the mutual inclination,
as well as planet X's mass and semimajor axis. The most successful simulations occur when the mutual inclination satisfies
either $50^\circ \lesssim i_{XB} \lesssim 70^\circ$ or $110^\circ \lesssim i_{XB} \lesssim 130^\circ$. For more aligned or anti-aligned configurations
($i_{XB} < 50^\circ$ or $>$130$^\circ$),
the obliquity of planet~b is not significantly excited,
and for near-polar configurations ($70^\circ < i_{XB} < 110^\circ$), eccentricity growth driven by von Zeipel--Kozai--Lidov oscillations provokes dynamical instability.

\begin{figure}
    \centering
    \includegraphics[width=\linewidth]{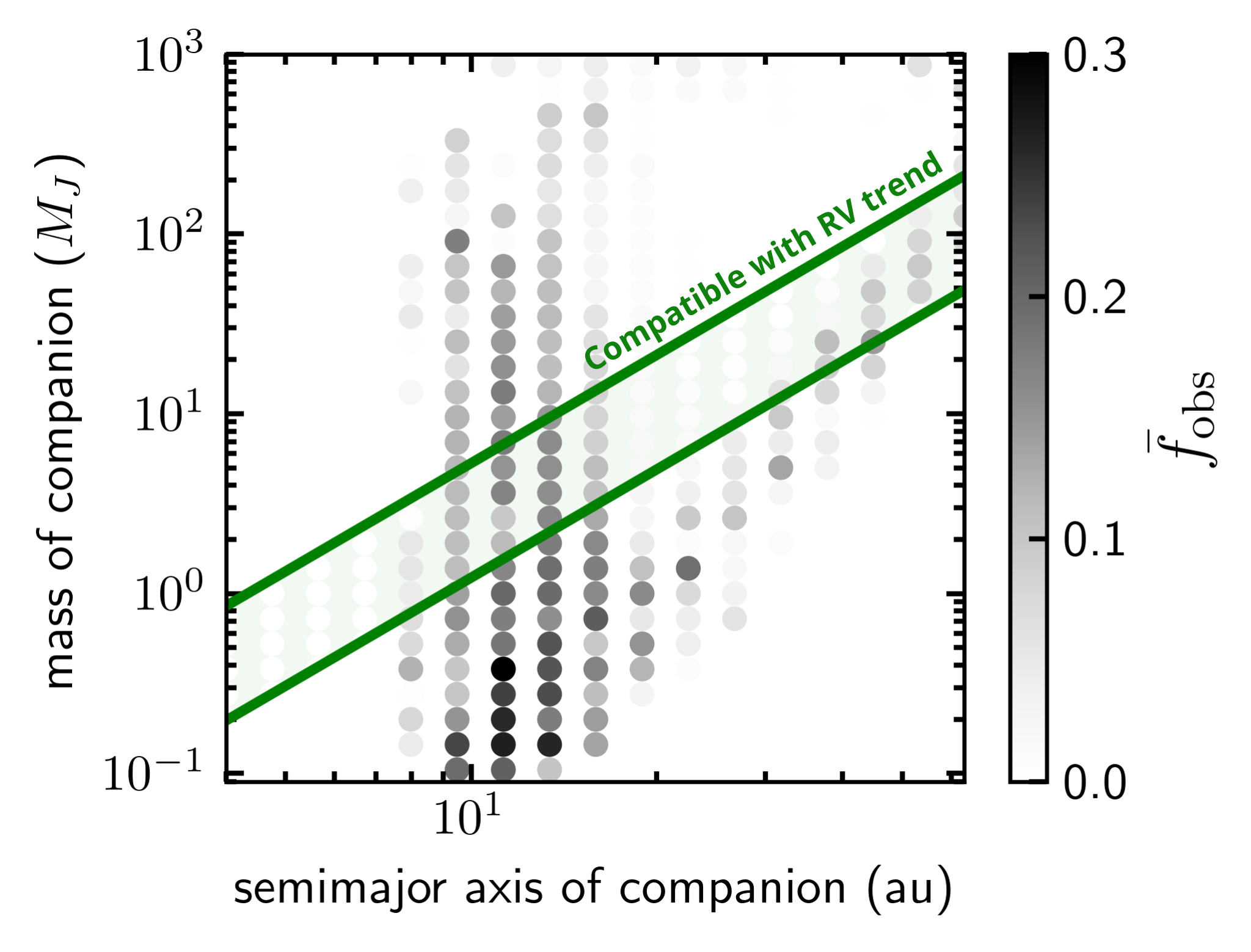}
    \caption{Mass--semimajor axis diagram for the hypothetical intermediate companion. The points are colored according to the probability of observing TOI-1710~A~b in its current configuration, as computed from our secular integrations. Each point represents an average over a range of initial mutual inclinations between the intermediate companion and the M-dwarf companion. The green region indicates the $5\sigma$ constraints imposed by the observed radial velocity trend.}
    \label{fig:companion}
\end{figure}

From this ensemble, we compute the inclination-averaged value of $f_{\rm obs}$ and identify the optimal parameters of planet~X as those that maximize $\bar{f}_{\rm obs}$. Figure~\ref{fig:companion} shows $\bar{f}_{\rm obs}$ as a function of the mass and semimajor axis of planet~X. These results indicate that planet~X is most likely to have a mass $\gtrsim 0.2\,M_{\rm J}$, with a preferred semimajor axis that depends on its mass. Within the green
boundaries, the companion's properties are broadly
compatible with the observed RV trend,
subject to uncertainties in the other orbital parameters.
The region of overlap between the optimal
parameters for orbit-flipping and the RV-based constraint
corresponds to a planet of $\sim 5\,M_{\rm J}$ at $\sim 15$~au.
A brown dwarf of $\sim 50\,M_{\rm J}$ at $\sim 50$~au would
also produce a substantial $\bar{f}_{\rm obs}$ and remains consistent with the observed radial-velocity trend. However, part of the parameter space in this region lies close to the stability boundary; indeed, several simulations with high initial $i_{XB}$ are found to be unstable according to the criteria of \citet{Mardling2001}.

In summary, our simulations successfully reproduce the observed properties of the inner planet, most notably its low eccentricity and retrograde orbit, over a broad
range of masses and semimajor axes of an intermediate companion. If this companion exists and is responsible for the observed RV
trend, it is most likely a Jovian planet with mass $\sim 5\,M_{\rm J}$, orbital separation $\sim 15$~au, and an orbit that is nearly aligned with that of the transiting planet. Future astrometric observations, particularly with Gaia, may be able to confirm the presence of this companion and provide further insight into the dynamical history of the system.

\begin{acknowledgments}

We thank Alex Polanski for kindly sharing the HIRES and APF RV observations of TOI-1710~A. \added{We would also like to thank the anonymous referee for their thoughtful review and suggestions that improved the quality of this work.}

\added{While this paper was under review, we learned of the independent RM measurement of TOI-1710~A~b by \cite{Mantovan2026}. The two stellar obliquity measurements are consistent with each other.}

This work is based on observations taken with the NEID instrument on the WIYN 3.5 m telescope at Kitt Peak National Observatory (Proposal ID 2025B-490658, PI Joshua Winn). We thank the NEID Queue Observers and WIYN Observing Associates for their skillful execution of our NEID observations. The authors are honored to be permitted to conduct astronomical research on I'oligam Du'ag (Kitt Peak), a mountain with particular significance to the Tohono O'odham. Kitt Peak is a facility of NSF's NOIRLab, managed by the Association of Universities for Research in Astronomy (AURA). The WIYN telescope is a joint facility of NOIRLab, Indiana University, the University of Wisconsin-Madison, Pennsylvania State University, Purdue University, and Princeton University. NEID was funded by the NASA-NSF Exoplanet Observational Research (NN-EXPLORE) partnership and built by Pennsylvania State University. The NEID archive is operated by the NASA Exoplanet Science Institute at the California Institute of Technology. NN-EXPLORE is managed by the Jet Propulsion Laboratory, California Institute of Technology under contract with the National Aeronautics and Space Administration. 

This paper includes data collected with the TESS mission, obtained from the MAST data archive at the Space Telescope Science Institute (STScI). Funding for the TESS mission is provided by the NASA Explorer Program. STScI is operated by the Association of Universities for Research in Astronomy, Inc., under NASA contract NAS 5–26555.

This research has made use of the Keck Observatory Archive (KOA), which is operated by the W.\ M.\ Keck Observatory and the NASA Exoplanet Science Institute (NExScI), under contract with the National Aeronautics and Space Administration.

J.N.W.\ acknowledges a grant associated with NEID observing programs from the NN-EXPLORE program (JPL RSA 1719095).

R.B.\ acknowledges support from Fondecyt Project 1241963.

A.J.\ acknowledges support from Fondecyt project 1251439.

\facilities{WIYN (NEID), TESS, TNG (HARPS-N), Keck:I (HIRES), OHP:1.93m (SOPHIE), APF (Levy), MAST.}

\software{
\texttt{astropy} \citep{astropy,astropy2,astropy3},
\texttt{batman} \citep{batman},
\texttt{celerite} \citep{celerite},
\texttt{ceres} \citep{ceres},
\texttt{dynesty} \citep{dynesty2},
\texttt{ironman} \citep{Espinoza-Retamal2023b,Espinoza-Retamal2024},
\texttt{juliet} \citep{juliet},
\texttt{lightkurve} \citep{lightkurve},
\texttt{lofti\_gaia} \citep{Pearce2020},
\texttt{matplotlib} \citep{matplotlib},
\texttt{numpy} \citep{numpy},
\texttt{radvel} \citep{Fulton18},
\texttt{rmfit} \citep{Stefansson2022},
\texttt{scipy} \citep{scipy},
\texttt{serval} \citep{Zechmeister2018},
\texttt{zaspe} \citep{zaspe}.
}

\end{acknowledgments}

\bibliography{sample7}{}
\bibliographystyle{aasjournalv7}

\appendix
\restartappendixnumbering

\section{NEID RV Measurements}\label{app:rvs}

Table \ref{tab:rv} presents the RVs of TOI-1710~A taken with NEID during the transit of the planet.

\begin{deluxetable}{ccc}[h!]
\digitalasset
\tablecaption{RV measurements from NEID.\label{tab:rv}}
\tablehead{
   \colhead{BJD} & \colhead{RV (m/s)} & \colhead{$\sigma_{\rm RV}$ (m/s)}
}
\startdata
2460973.74158 & 0.4 & 1.2 \\
2460973.748957 & 0.1 & 1.3 \\
2460973.756173 & -3.2 & 1.4 \\
\nodata & \nodata & \nodata \\
2460974.00989 & 2.5 & 0.9 \\
2460974.017167 & 2.2 & 0.9 \\
2460974.024408 & 1.6 & 0.9 \\
\enddata
\tablecomments{This table is available in its entirety in machine-readable form in the online article. A portion is shown here for guidance regarding its form and content.}
\end{deluxetable}

\clearpage

\section{Stellar Parameters}\label{app:stellar}

Table \ref{tab:stellar} presents the stellar parameters of TOI-1710~A derived as described in Section \ref{sec:stellar}.

\begin{deluxetable*}{llcr}[h!]
\tablecaption{Stellar properties of TOI-1710 A.\label{tab:stellar}}
\tablecolumns{4}
\tablewidth{0pt}
\tablehead{Parameter & Description & TOI-1710 A & Reference}
\startdata
RA & Right Ascension (J2015.5) & 06h17m08.12s & \citet{GaiaDR3}\\
Dec & Declination (J2015.5) & 76d12m39.67s & \citet{GaiaDR3}\\
pm$^{\rm RA}$ & Proper motion in RA (mas yr$^{-1}$)  & 59.64$\pm$0.01 & \citet{GaiaDR3}\\
pm$^{\rm Dec}$ & Proper motion in DEC (mas yr$^{-1}$) & 55.668$\pm$0.011 & \citet{GaiaDR3}\\
$\pi$ & Parallax (mas) & 12.325$\pm$0.010 & \citet{GaiaDR3} \\
$d$ & Distance (pc) & 81.2$\pm$0.1 & \citet{GaiaDR3} \\
\hline
T & TESS magnitude (mag) & 8.913$\pm$0.006 & \citet{Stassun2018,Stassun2019}\\
B  & B-band magnitude (mag) & 10.20$\pm$0.04 & \citet{apass}\\
V  & V-band magnitude (mag) & 9.545$\pm$0.003 & \citet{apass}\\
G  & Gaia G-band magnitude (mag) & 9.3674$\pm$0.0001 & \citet{GaiaDR3}\\
G$_{\rm BP}$ & Gaia BP-band magnitude (mag) & 9.7055$\pm$0.0003 & \citet{GaiaDR3}\\
G$_{\rm RP}$ & Gaia RP-band magnitude (mag) & 8.8600$\pm$0.0003 & \citet{GaiaDR3}\\
J & 2MASS J-band magnitude (mag) & 8.319$\pm$0.019 & \citet{2mass}\\
H & 2MASS H-band magnitude (mag) & 8.003$\pm$0.034 & \citet{2mass}\\
K$_s$ & 2MASS K$_s$-band magnitude (mag) & 7.959$\pm$0.026 & \citet{2mass}\\
\hline
$T_{\rm eff}$ & Effective temperature (K) & 5775$\pm$80 & This work\\
$\log{g}$ & Surface gravity (cgs) & 4.51$\pm$0.01 & This work\\
$[$Fe/H$]$ & Metallicity (dex) & +0.04$\pm$0.05 & This work\\
$v\sin{i_\star}$ & Projected rotational velocity (km s$^{-1}$) & 2.1$\pm$0.3 & This work\\
$M_{\star}$ & Mass ($M_\odot$) & 1.023$\pm$0.015 & This work\\
$R_{\star}$ & Radius ($R_\odot$) & 0.93$\pm$0.01 & This work\\
$L_{\star}$ & Luminosity ($L_\odot$) & $0.86_{-0.01}^{+0.02}$ & This work\\
$A_{V}$ & Visual extinction (mag) & $0.027_{-0.015}^{+0.025}$ & This work\\
Age & Age (Gyr) & $1.0_{-0.6}^{+1.1}$ & This work\\
$\rho_\star$ & Mean density (g cm$^{-3}$) & $1.80_{-0.07}^{+0.04}$ & This work\\
$P_{\rm rot}$ & Rotational period (d) & $22.5\pm2.0$ & \citet{Konig2022}\\
\enddata
\tablecomments{The stellar parameters computed in this work do not consider possible systematic differences among different stellar evolutionary models and have underestimated uncertainties \citep{Tayar2022}. The TESS magnitude is shown only for reference and was not included in our stellar analysis.}
\end{deluxetable*}

\clearpage

\section{Secular four-body dynamics}\label{app:simulations}

We model a four-body system consisting of the host star (A), the known planet (b), an intermediate object (X), and a distant M-dwarf companion (B). Treating the system as a hierarchical configuration, we use secular equations of motion to track its long-term evolution. 

The secular Hamiltonian for a 3+1 hierarchical system was derived by \citet{hamersSecularDynamicsHierarchical2015}, who expanded the Hamiltonian in the ratio of semimajor axes of the interacting orbits up to hexadecapole order. Here, we adapt and simplify their formulation for our specific model. To track the long-term evolution of planet~b, we include octupole-order perturbations from planet~X. For planet~X, we include octupole-order perturbations from both planet~b and the stellar companion~B. For companion~B, however, we retain only quadrupole-order perturbations from planet~X. Octupole-order perturbations from the intermediate companion scale as $e_X\, e_B (a_X/a_B)^{3/2}/(1 - e_B^2)$, and since in our simulations $e_X \approx 0$ and $a_B \gg a_X$, these terms can be safely neglected. We also neglect cross terms in the Hamiltonian expansion arising from the direct perturbations of companion~B on planet~b, as their characteristic von Zeipel-Kozai-Lidov (ZKL) timescale, $t_{\rm ZKL,bB} \sim 400\,\mathrm{Gyr}$, is far longer than the system’s age.

For the orbit of planet b, we include short-range forces---general relativistic precession, rotational and tidal deformation---and model tidal dissipation using the equilibrium tide framework. We adopt viscous timescales of $t_{v,A} = 50\,\mathrm{yr}$ for the star, and $t_{v,b} = 0.1\,\mathrm{yr}$ for the planet, with Love numbers $k_{2,A} = 0.028$ and $k_{2,b} = 0.5$, respectively. The initial stellar and planetary spin periods are set to 22 days and 0.5 days. See \cite{Petrovich2015} for the relevant equations of motion.

For the inner triple, long-term stability
was checked using the stability criterion from \cite{PetrovichStab2015}, which is appropriate for planetary systems. For the outer triple—comprising the host star, the intermediate companion, and the M‑dwarf—we use the criterion from \cite{Mardling2001}. In our simulations, instability arises primarily within the inner triple. This mainly occurs when the M‑dwarf’s orbital plane is inclined by more than $\sim60^\circ$ relative to the inner system, and the intermediate companion has a relatively wide orbit ($>20$ AU). Such a configuration induces eccentricity oscillations in the intermediate companion, which subsequently drives it close to the observed planet, ultimately destabilizing the system. In our analysis, we ignore simulations which have been marked as unstable.

\begin{figure}
    \centering
    \includegraphics[width=\linewidth]{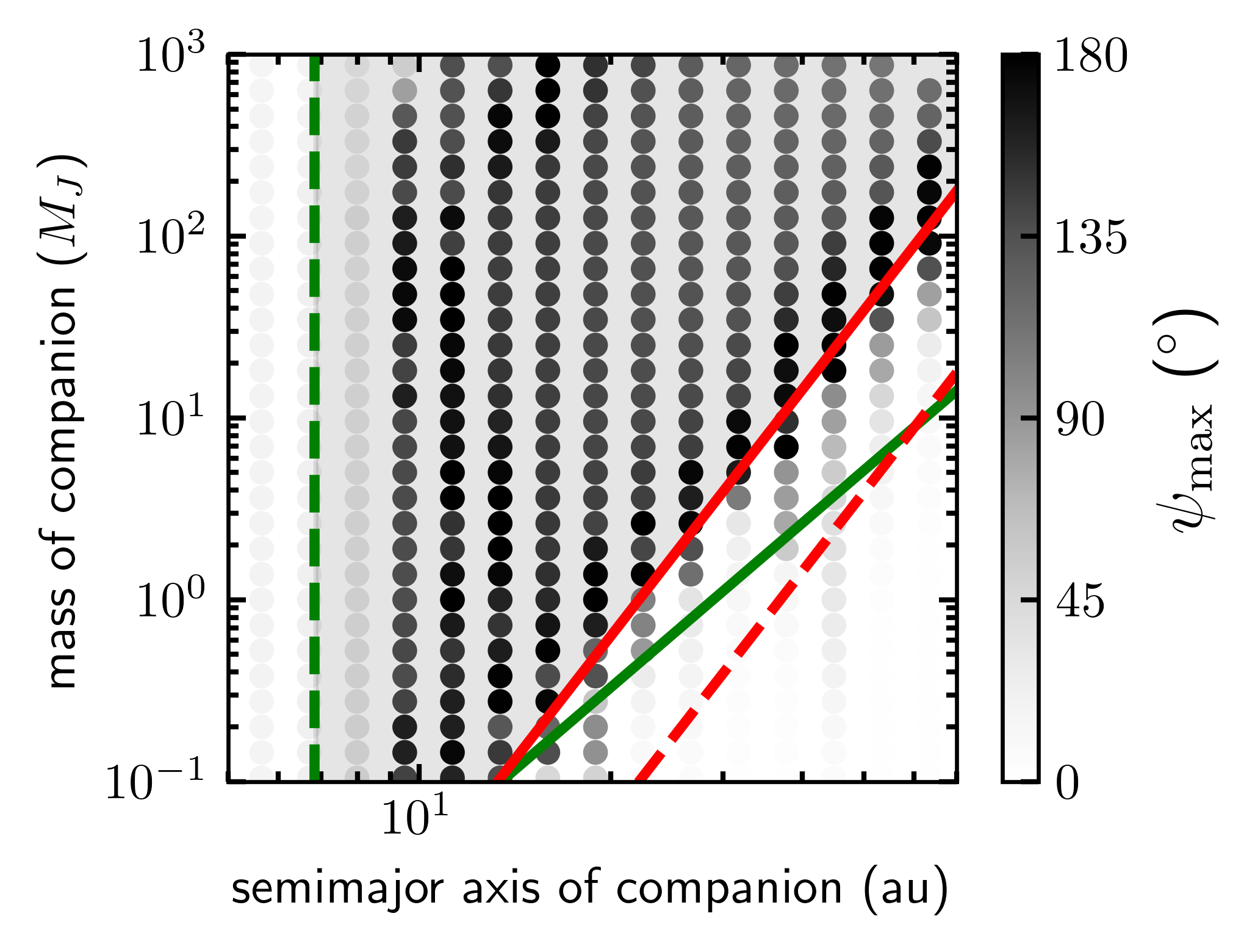}
    \caption{Results of the ensemble of secular simulations. The semimajor axis of the intermediate companion X is shown on the x-axis, and its mass is shown on the y-axis. The color scale indicates the maximum obliquity attained by the observed planet b. In all simulations, the initial inclination between the planets and the M-dwarf companion is set to $50^\circ$. Analytical expressions delineating the different dynamical regimes are shown as solid and dashed lines. See the text for a detailed explanation. We use the following initial conditions to run these simulations: $a_b=0.165$ AU, $e_b=0.05$, $e_X=0.1$, $i_{bX}=2.6^\circ$, and $i_{XB}=56^\circ$. The initial spin periods of the host star and the observed planet are taken to be 22 days, and 0.5 days respectively. The initial obliquities of the planets are set close to zero ($=0.025$). The viscous dissipation timescales of the star and the planet are set to 50 years and 0.1 years, respectively. Meanwhile, the love numbers of the star and the observed planet are set to $k_{2,0}$= 0.028 and $k_{2,p}=$ 0.5, respectively.}
    \label{fig:parspace_epsimax}
\end{figure}

Figure \ref{fig:parspace_epsimax} shows results from our ensemble of simulations. When planet X orbits close to the host star ($a_X < 7\,\mathrm{au}$), both the eccentricity and obliquity of planet b remain low ($e_{b,\max} \sim 0.05$, $\psi_{b,\max} < 10^\circ$). This region is bounded on the right side by the \textit{dashed green line}, which marks the semimajor axis at which the precession timescale of the intermediate companion due to the M-dwarf companion is comparable to the age of the system ($\sim 10^9$  years). This timescale is given by the ZKL timescale: 
\begin{equation}
t_{\mathrm{\rm ZKL,XB}} \equiv 
\left(\frac{M_{\mathrm{A}}}{M_{\mathrm{B}}}\right)
\left(\frac{a_{\mathrm{B}}}{a_{\mathrm{X}}}\right)^3
\left( 1 - e_{\mathrm{B}}^{2} \right)^{3/2} P_{X},
\end{equation}
\citep[e.g.,][]{Antogini2015}. To the left of this line, planet X is effectively decoupled from the M-dwarf. In this regime, the star's obliquity relative to planet b remains low and incompatible with observations.

A second low-excitation regime exists where the intermediate companion is both low-mass ($M_X \lesssim 10^{-2}$) and distant ($a_X \gtrsim 20\,\mathrm{AU}$). This region is bounded on the left side by the \textit{solid green line}, where the semimajor axis at which the nodal precession rate of planet b induced by the stellar $J_2$ becomes comparable to the nodal precession rate induced by planet X. This condition
is obtained by using the expression for Laplace radius of
planet b:
\begin{equation}
a_b = r_L = \left( J_{2} R_A^2 a_{\mathrm{X}}^3 
\frac{M_A}{M_{\mathrm{X}}}\right)^{1/5}.
\end{equation}
Here, $R_A$ is the radius of the host star \citep[e.g., ][]{Tremaine2009} and $J_2$ is the star's gravitational moment:
\begin{equation}
J_2=\frac{k_{2,A}}{3}\left(\frac{\Omega_{s,A}}{\Omega_{s,brk}}\right)^2,
\end{equation}
where $\Omega_{s,A}$ is the spin rate of the host star and $\Omega_{s,brk}=\sqrt{GM_A/R_A^3}$ is the breakup rate. To the right of the solid green line, the stellar $J_2$ dominates and the influence of planet X is negligible. This region is inconsistent with observations, as the obliquity remains unexcited.

The elevated obliquities near $a_X \sim 10\,\mathrm{AU}$ arise from a spin--orbit resonance. This occurs when the stellar spin precession rate (driven by torques from the inner planet) becomes comparable to the nodal precession rate induced by the M-dwarf companion on planet X. In this regime, the intermediate companion and planet b remain strongly coupled and precess together.

The stellar spin precession rate induced by planet b is given by
\[
\alpha = \frac{k_{A}}{k_{I,A}} \frac{M_b}{M_A} \left( \frac{R_A}{a_b} \right)^3 \Omega_{s,A},
\]
where $k_A = k_{2,A}/2$ is the stellar apsidal motion constant and $k_{I,A} = 0.08$ is the moment-of-inertia constant. For this system, we obtain $\alpha = 1.8 \times 10^{-8}\,\mathrm{yr}^{-1}$. A spin--orbit resonance occurs when this rate becomes comparable to the nodal precession rate (i.e., $\alpha t_{\rm ZKL,XB} \sim 1$). In this regime, the stellar spin undergoes large secular variations, allowing high obliquities to be reached.

Another distinct region where the obliquity of the observed planet is strongly excited ($\psi_{b,\max} > 50^\circ$)
involves a different secular orbital resonance:
the nodal precession rate of planet X due to the M-dwarf matches the precession rate of planet b due to planet X. The contours of the precession-rate ratio $\mathcal{R} = 1$ and $\mathcal{R} = 10$ are shown as solid and dashed red lines, respectively. The analytical estimate for $\mathcal{R}$ is given by \citep{Hamers2017}:
\begin{eqnarray}
\mathcal{R}&\equiv&\frac{t_{\mathrm{ZKL},bX}}{t_{\mathrm{ZKL},XB}} \nonumber \\
&\simeq& \frac{M_B}{M_X}a_b^{-\frac{3}{2}}a_X^{\frac{9}{2}}a_B^{-3}\frac{1}{(1-e_B^2)^{\frac{3}{2}}}.
\end{eqnarray}
This condition marks the onset of chaotic behavior, which can drive large obliquity excitation as recently shown by Liu et al. (in prep.).

Overall, there is a broad region of parameter space in which the obliquity of the observed planet can be excited to large values, consistent with the observations.

\end{document}